\RequirePackage{fix-cm}
\documentclass[smallextended]{svjour3}       
\smartqed  
\usepackage{graphicx}
\usepackage{epstopdf}
\usepackage{color}
\usepackage{cite}
\usepackage{amsmath}
\journalname{Quantum Information Processing}
\begin{document}

\title{CNOT gate on reverse photon modes in ring cavity
\thanks{This work was funded by RFBR according to the research project no. 17-02-00918.}
}


\author{Sergey N. Andrianov         \and
        Narkis M. Arslanov \and
        Konstantin I. Gerasimov \and
        Alexander A. Kalinkin \and
        Sergey A. Moiseev
}


\institute{Sergey N. Andrianov \at
              Institute for Applied Research of Tatarstan Academy of Sciences, 36a Levobulachnaya street, Kazan, Russian Federation 422111 \\
              \email{andrianovsn@mail.ru}           
           \and
           Narkis M. Arslanov \at
              Kazan Quantum Center of Kazan National Research Technical University, 10 K. Marx street, Kazan, Russian Federation 420111 \\
            \email{narkis@yandex.ru}
           \and
             Konstantin I. Gerasimov \at
              Kazan Quantum Center of Kazan National Research Technical University, 10 K. Marx street, Kazan, Russian Federation 420111 \\
            \email{kigerasimov@mail.ru}
           \and
             Alexander A. Kalinkin \at
              M.V. Lomonosov Moscow State University, Faculty of Physics, 1-2 Leninskie Gory, Moscow, Russian Federation 119991 \\
              and Quantum Technology Centre of  MSU, 119991, Moscow, 1-2 Leninskie Gory, Moscow, Russian Federation 119991 \\
            \email{kalinkin@physics.msu.ru}
           \and
             Sergey A. Moiseev \at
              Kazan Quantum Center of Kazan National Research Technical University, 10 K. Marx street, Kazan, Russian Federation 420111 \\
            \email{samoi@yandex.ru}
}

\date{Received: date / Accepted: date}

\maketitle

\begin{abstract}
Photon modes of reverse rotation in ring QED-cavity coupled with single atom are considered. By applying Schrieffer-Wolf transformation for the off-resonant light-atom interaction, an effective Hamiltonian of the evolution of the photon modes is obtained. Heisenberg equations for the input-output photon mode operators are written and expression for the system wave function is found. The obtained analytical solution shows the condition of the control not quantum gate implementation. Possible on a chip experimental implementation and recommendations for the construction of optical quantum computer using this gate are considered.
\end{abstract}

\section{Introduction}
\label{intro}
Photonic two-qubit gates are important instruments for quantum processing, optical quantum communication etc. However its effective experimental realization remains a quite difficult problem. Since the photons do not interact with each other in free space, the photonic two qubit gate are realized in probabilistic way using linear optical elements \cite{Knill_2001}. It is possible to enhance the photon –photon interaction in nonlinear media. In particular, the authors of paper \cite{Niu_2018} have proposed using additional nonlinear two-photon frequency up and down conversion elements. This nonlinear interaction is also rather weak but can be considerably enhanced in micro- and nano-cavities due to the Purcell effect. Moreover, this indirect interaction can be replaced by the resonant interaction of photons with atoms. Therefore, the theme of construction photon-photon two-qubit gates using cavities containing atoms is a subject of intensive investigation.

CPhase photon-photon gate using the interaction of polarization encoded photon qubits with a cavity containing three-level atom was presented in paper \cite{Duan_2004}. CPhase gate on nitrogen vacancy center in this configuration was considered in papers \cite{Wang_2013, Ren_2017}. Three qubit control SWAP (Toffoli) gate was proposed in paper \cite{Wang_2007}.
Deterministic photon-photon $\sqrt{SWAP}$ gate using a $\Lambda$-type atom-cavity system with polarization encoding of qubits was theoretically proposed in paper \cite{Koshino_2010}.
Also, CPhase photon-photon gate in this approach was proposed in paper \cite{Tokunaga:15}.
CNOT gates were proposed for photon qubit interacting with cavity containing atom \cite{Wang_2016}, quantum dot \cite{Wei_2013, Wei_OE_2013}, diamond nitrogen-vacancy centers \cite{Wei_2015} and collective magnon system \cite{Liu_2017}.
Recently, the realization of original scheme \cite{Duan_2004} was successfully demonstrated in experiment \cite{Hacker_2016}.
Experimental realization of SWAP gate using nanofiber-coupled microsphere resonator coupled to single Rb atom was presented in paper \cite{Kim_2013}. CNOT gate on quantum dot in photonic crystal cavity strongly coupled to polarization encoded photon qubits was demonstrated in paper \cite{Schrieffer_1966}.

Thus, large number of theoretical and experimental works concerning photon-photon cavity based quantum gates has been performed, although the quantum efficiency and fidelity of the two-qubit gates remain insufficient high \cite{Hacker_2016}, which requires further searching for new ways to improve practical realization of two-photon gates.
Besides, aforementioned approaches used polarization encoding of photon qubits and it is interesting to study the possibility to use other encodings.
Here, we propose using the dispersion interaction of light modes with the atom in order to suppress influence of atomic decoherence processes.
Moreover, here we propose  photon-photon CNOT gate using encoding of qubits on contrary directed photon modes of propagation in ring cavity with controlling two-level atom and consider the on a chip scheme for its experimental implementation.

\section{Schrieffer-Wolf transformation and effective Hamiltonian}

\begin{figure}
  \includegraphics[width=1\textwidth]{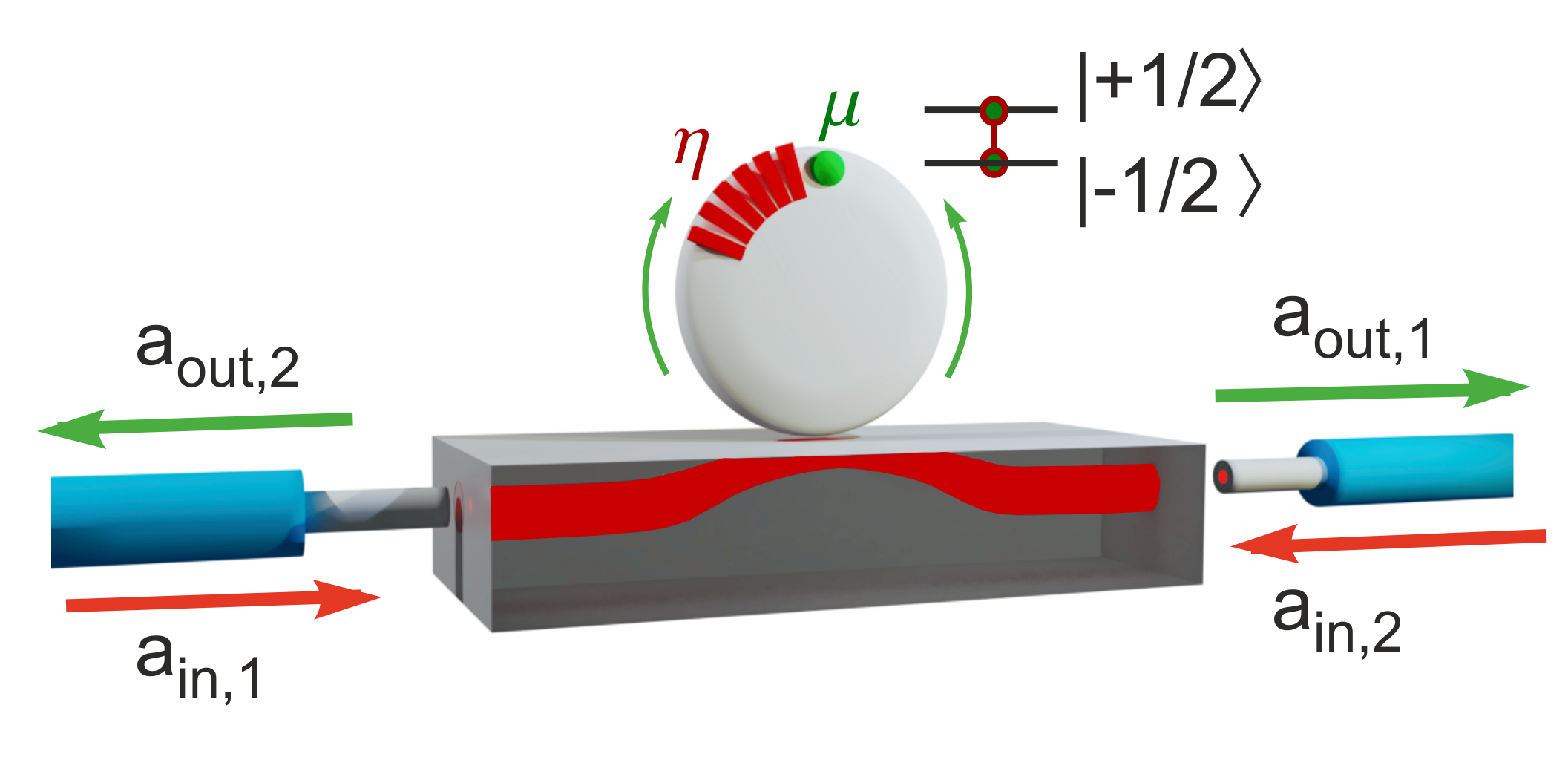}
\caption{On a chip scheme  for photon-photon CNOT gate scheme;   reverse whispering-gallery modes of optical  cavity interacts with two-level atom with effective coupling constant $\mu$ and  with each other with coupling constant $\eta$  by scattering in Bragg-grating situated closely to the cavity surface.  }
\label{fig:1}       
\end{figure}

The analyzed scheme is depicted in Fig.\ref{fig:1}  where resonant atom is placed in the ring cavity connected via waveguide with multi-qubit quantum memory cell. The ring cavity supplies two counter propagating light modes coupled with each other due to scattering on the tuned Bragg-grating. We assume that the atom interacts with two photons oppositely rotating in the ring cavity. The total Hamiltonian of studying system is written as follows $H=H_0+H_1$, where $H_0=H_a+H_r$ is main Hamiltonian and $H_1=H^{\left(cw\right)}_{r-a}+H^{\left(ccw\right)}_{r-a}$ is perturbation Hamiltonian. In this scheme, we focus to considering only a two-level atom, but the obtained results can be also applied to the case of a three-level atom, where the specific features and advantages arising from this will be briefly discussed below.\textit{ }Main Hamiltonian (in units $\hbar $) consists of atomic Hamiltonian $H_a={\omega }_0S^z$, where ${\omega }_0$ is frequency of atomic transition in two-level model, $S^z$ is operator of effective spin z-component (i.e. we ignore decoherent processes in the atomic dynamics), and radiation Hamiltonian $H_r={\omega }_{cw}a^{\dagger}_{cw}a_{cw}+{\omega }_{ccw}a^{\dagger}_{ccw}a_{ccw}+\eta (a^{\dagger}_{cw}a_{ccw}+a^{\dagger}_{ccw}a_{cw})$, where ${\omega }_{cw}$ and ${\omega }_{ccw}$ are frequencies of clockwise and counterclockwise photons in ring cavity, $a^+_{cw\left(ccw\right)}$ and $a_{cw\left(ccw\right)}$ are creation and annihilation operators for clockwise (counterclockwise) photon, $\eta $ is clockwise-counterclockwise photons interaction constant. The coupling of the counterpropagating modes can be realized by forming Bragg-grating with parameters (period and amplitude) determining appropriate value $\lambda $.

Perturbation Hamiltonian includes atomic interaction with clockwise and counterclockwise photons $H^{\left(cw\right)}_{r-a}=gS^+a_{cw}+g^*S^-a^{\dagger}_{cw}{\rm \ }$ and similarly $H^{\left(ccw\right)}_{r-a}=gS^+a_{ccw}+g^*S^-a^{\dagger}_{ccw}$, where $S^+$ and $S^-$ are raising and lowering operators in two-level model, \textit{g} is interaction constant. Below we consider off-resonant interaction between light modes and two-level atom by assuming some detuning $\Delta $ of the atomic frequency ${\omega }_0$ from the cavity mode frequency ${\omega }_{ccw}$. In this case we get dispersion light-atom interaction where atomic decoherent processes are suppressed by the factor $\frac{g}{\Delta }\ll {\rm 1}$ \cite{Schrieffer_1966}. Performing Shrieffer-Wolf transformation \cite{Seri_2018}  $e^s$ for this case, we get for effective Hamiltonian $H_s=H_0+\frac{1}{2}\left[H_1,s\right]$\textit{ }providing $H_1+\left[H_0,s\right]=0$ that gives the following formula for transformation operator:

\begin{equation}
\label{eq1}
 s={\alpha }^{\left(cw\right)}gS^+a_{cw}+{\beta }^{\left(cw\right)}g^*S^-a^{\dagger}_{cw}+{\alpha }^{\left(ccw\right)}gS^+a_{ccw}+{\beta }^{\left(ccw\right)}g^*S^-a^{\dagger}_{ccw},
\end{equation}

\noindent
where ${\Delta =\omega }_0-{\omega }_{cw},\ {\alpha }^{\left(cw\right)}=-{\beta }^{\left(cw\right)}=-\frac{1}{\Delta -\eta }$ , ${\alpha }^{\left(ccw\right)}=-{\beta }^{\left(ccw\right)}=-\frac{1}{\Delta -\eta }$ and it is assumed ${\beta }^{\left(ccw\right)}g\ll 1$, ${\beta }^{\left(cw\right)}g\ll 1$ and $\Delta \gg $ determines dispersive interaction of light with atom,  is a decay constant of the atomic coherence, ${\omega }_{cw}={\omega }_{ccw}={\omega }_r$, ${\beta }^{\left(cw\right)}{=\beta }^{\left(ccw\right)}=\beta $.

By using (\ref{eq1}) we get the effective Hamiltonian:

\begin{equation}
\label{eq2}
\begin{array}{ll}
   H_s=&{\omega }_{cw}a^{\dagger}_{cw}a_{cw}+{\omega }_{ccw}a^{\dagger}_{ccw}a_{ccw}+\eta \left(a^{\dagger}_{cw}a_{ccw}+a^{\dagger}_{ccw}a_{cw}\right)+ \left({\omega}_0+2\beta {\left|g\right|}^2\right)S^z
   \nonumber  \\
   &+  \mu S^z\left({a^{\dagger}_{cw}a_{cw}+a^{\dagger}_{ccw}a_{ccw}+a}_{cw}a^{\dagger}_{ccw}+a^{\dagger}_{cw}a_{ccw}\right),
\end{array}
\end{equation}

\noindent
where the coupling between the cavity modes is controlled by the atomic state via $\mu S^z$ (where $\mu =2\beta {\left|g\right|}^2=2{\left|g\right|}^2/(\ \Delta -\eta )$) with $S^z=\pm 1/2$.
Below we analyze the dynamics of the field modes.

\section{System dynamics and CNOT gate}

Now, accounting coupling of cavity with an external waveguide and using the well-known approach of quantum optics \cite{Kim_2013}, we obtain two coupled equations:

\begin{equation}
\label{eq3}
{\frac{d}{dt}a_{cw}}=-\left(i{\omega }_{cw}+\frac{\kappa }{2}\right)a_{cw}-i\left\{\eta +\mu S^z\right\}a_{ccw}+\sqrt{\kappa }\ a_{in,1}\left(t\right), \end{equation}

\begin{equation}
\label{eq4}
{\frac{d}{dt}a_{ccw}}=-i({\omega }_{ccw}+\frac{\kappa }{2})a_{ccw}-i\{\eta +\mu S^z\}a_{cw}+\sqrt{\kappa }\ a_{in,2}(t),
\end{equation}

\noindent
where $\kappa $ is coupling constant of the cavity mode with free modes, $a_{in,(1,2)}$ is the input field of m-cavity mode launched from the quantum memory cell (see Fig.1).

Linear equations for field operators $a_{cw}(t)$ and $a_{ccw}(t)$ are easily solved for its Fourier components and yield for two states of atom ($m=\pm 1/2$):

\begin{equation}
    \label{eq5}
    a_{cw}\left(\omega \right)=\frac{-i\{\eta +\mu m\}\sqrt{\kappa }\ a_{in,2}\left(\omega \right)+\left(\frac{\kappa }{2}-i\left(\omega -{\omega }_{ccw}\right)\right)\sqrt{\kappa }\ a_{in,1}\left(\omega \right)}{{\left(\eta +\mu m\right)}^2+\left(\frac{\kappa }{2}-i\left(\omega -{\omega }_{cw}\right)\right)\left(\frac{\kappa }{2}-i\left(\omega -{\omega }_{ccw}\right)\right)},
\end{equation}

\begin{equation}
    \label{eq6}
   a_{ccw}\left(\omega \right)=\frac{-i\left\{\eta +\mu m\right\}\sqrt{\kappa }\ a_{in,1}\left(\omega \right)+\left(\frac{\kappa }{2}-i\left(\omega -{\omega }_{cw}\right)\right)\sqrt{\kappa }\ a_{in,2}\left(\omega \right)}{{\left(\eta +\mu m\right)}^2+\left(\frac{\kappa }{2}-i\left(\omega -{\omega }_{cw}\right)\right)\left(\frac{\kappa }{2}-i\left(\omega -{\omega }_{ccw}\right)\right)}.
\end{equation}

\noindent
Using these solutions and accounting the initial state for input light fields $a_{in,1}(t)$ and $a_{in,2}(t)$, we can get now features of output fields $a_{out,1}\left(t\right)\ $ and $a_{out,2}\left(t\right)$ from relations \cite{Walls_1994} $a_{out,1}\left(t\right)=\sqrt{\kappa }a_{cw}\left(t\right)-a_{in,1}(t)$, $a_{out,2}=\sqrt{\kappa }a_{ccw}-a_{in,2}$.
By assuming ${\omega }_{cw}={\omega }_{ccw}={\omega }_r$, we get from (\ref{eq5}, \ref{eq6}):

\begin{equation}
    \label{eq7}
    a_{out,1}\left(\omega \right)=\ B_{1,1}(m,\omega )a_{in,1}\left(\omega \right)+{B_{1,2}(m,\omega )a}_{in,2}\left(\omega \right),
\end{equation}
\begin{equation}
    \label{eq8}
a_{out,2}\left(\omega \right)=B_{2,1}(m,\omega) a_{in,1}\left(\omega \right)+B_{2,2}(m,\omega) a_{in,2}\left(\omega \right),
\end{equation}

\noindent
where
\begin{equation*}
    \label{eq8a}
B_{1,2}(m,\omega )=B_{2,1}(m,\omega )=\frac{-i\left\{\eta +\mu m\right\}\kappa \ }{{\left(\eta +\mu m\right)}^2+{\left(\frac{\kappa }{2}-i\left(\omega -{\omega }_r\right)\right)}^2},
\end{equation*}
and
\begin{equation*}
    \label{eq8b}
{B_{1,1}(m,\omega )=B}_{2,2}(m,\omega )=-\frac{\left\{{\left(\eta +\mu m\right)}^2-\frac{{\kappa }^2}{4}-{\left(\omega -{\omega }_r\right)}^2\right\}}{{\left(\eta +\mu m\right)}^2+{\left(\frac{\kappa }{2}-i\left(\omega -{\omega }_r\right)\right)}^2}.
\end{equation*}

Eqs. \ref{eq7},\ref{eq8} demonstrate dependence of output fields on the state of the atom.
For the input single photon state

\begin{equation}
    \label{eq9}
         \begin{array}{cc}
    |{\Psi }_{in}\rangle= \\
    \sum^{-\frac{1}{2},+\frac{1}{2}}_m{{\alpha }_m\int^{\infty }_{-\infty }{{d\omega} \{C^{\left(m\right)}_1\left(\omega \right)a^{\dagger}_{in,1}\left(\omega \right)+}{C^{\left(m\right)}_2\left(\omega \right)}a^{\dagger}_{in,2}\left(\omega \right)\}{\left.|0\right\rangle }_1{\left.|0\right\rangle }_2{\left.|m\right\rangle }_c},
    \end{array}
\end{equation}
we get the output quantum state of light after the photon-atom interaction:

\begin{equation}
    \label{eq10}
     \begin{array}{cc}
|{\psi }_{at+ph}\rangle =& \nonumber \\
\sum^{-\frac{1}{2},+\frac{1}{2}}_m
\alpha_m &\int^{\infty }_{-\infty }
{d\omega \{C^{\left(m\right)}_{1,out}(\omega)a^{(m){\dagger}}_{out,1}(\omega)+
C^{(m)}_{2,out}(\omega)}a^{(m){\dagger}}_{out,2}\}
{\left.|0\right\rangle }_1
{\left.|0\right\rangle }_2
{\left.|m\right\rangle }_c.
\end{array}
\end{equation}
where

\begin{equation}
    \label{eq11}
    {C^{\left(m\right)}_{1,out}\left(\omega \right)}{a}^{\left(m\right){\dagger}}_{out,1}\left(\omega \right)=C^{\left(m\right)}_1\left(\omega \right)\ B^*_{1,1}{(m)}{a}^{\dagger}_{in,1}\left(\omega \right)+C^{\left(m\right)}_2\left(\omega \right)B^*_{1,2}{(m)}{a}^{\dagger}_{in,2}(\omega ),
\end{equation}

\begin{equation}
\label{eq12} {C^{\left(m\right)}_{2,out}\left(\omega \right)a}^{\left(m\right)+}_{out,2}\left(\omega \right)=C^{\left(m\right)}_1\left(\omega \right)B^*_{2,1}\left(m\right)a^+_{in,1}\left(\omega \right)+C^{\left(m\right)}_2\left(\omega \right)B^*_{2,2}\ \left(m\right)a^+_{in,2}\left(\omega \right).
\end{equation}

Below we assume the input photon state $C^m_{1.2}\left(\omega \right)$ is characterized by sufficiently narrow spectral width  $\frac{{\delta \omega }_s}{\kappa }\ll 1$, where $B^*_{1,2}(m)=B^*_{2,1}(m)=\frac{i\left\{\eta +\mu m\right\}\kappa \ }{{\left(\eta +\mu m\right)}^2+\frac{{\kappa }^2}{4}}$,
${B^*_{1,1}=B}^*_{2,2}=-\frac{\left\{{\left(\eta +\mu m\right)}^2-\frac{{\kappa }^2}{4}\right\}}{{\left(\eta +\mu m\right)}^2+\frac{{\kappa }^2}{4}}$.
By taking into account the fulfillment of the impedance matching conditions:
\begin{equation}
    \label{eq13}
\eta =\frac{\mu }{2},
\mu =\frac{\kappa }{2},
\end{equation}

\noindent
we find for the field amplitudes: $B^*_{1,2}(-1/2)=B^*_{2,1}(-1/2)=0$, ${B^*_{1,1}=B}^*_{2,2}=1\ \ $ if m=-1/2, and $B^*_{1,2}\left(\frac{1}{2}\right)=B^*_{2,1}\left(\frac{1}{2}\right)=i$; ${B^*_{1,1}=B}^*_{2,2}=0$ if m=1/2.
Finally, the wave function of the output fields will be

\begin{equation}
    \label{eq14}
    \begin{array}{cc}
 |{\psi }_{at+ph}\rangle=\\ \nonumber
 {\alpha }_{-1/2}\int^{\infty }_{-\infty }{d\omega \{C^{\left(-\frac{1}{2}\right)}_1\left(\omega \right)a^{\dagger}_{in,1}\left(\omega \right)+C^{\left(-\frac{1}{2}\right)}_2}a^{\dagger}_{in,2}\left(\omega \right){\}\left.|0\right\rangle }_1{\left.|0\right\rangle }_2{\left.|-1/2\right\rangle }_c  \\
+i{\alpha }_{1/2}\int^{\infty }_{-\infty }{d\omega \{C^{\left(\frac{1}{2}\right)}_1\left(\omega \right)a^{\dagger}_{in,2}\left(\omega \right)+C^{\left(\frac{1}{2}\right)}_2\left(\omega \right)}a^{\dagger}_{in,1}\left(\omega \right){\}\left.|0\right\rangle }_1{\left.|0\right\rangle }_2{\left.|1/2\right\rangle }_c.
    \end{array}
\end{equation}

\noindent
We note that the coupling constant $\eta $ can be varied by changing contrast of Bragg-grating and its distance to the surface of ring cavity as it was used recently for nanofiber cavity containing one atom \cite{Yalla_2014} (see Fig.\ref{fig:1})

By performing edition rotation of the atom state on angle $\pi/2$ we get :

\begin{equation}
        \begin{array}{ll}
|{\Psi }_{out}\rangle={\exp  \left\{-\frac{iS_z\pi }{2}\right\}\ }|{\psi }_{at+ph}\rangle= \exp  \left\{i\pi/4\right\}\cdot  \\
\{\alpha_{-1/2}\int^{\infty}_{-\infty }{d\omega[C^{\left(-\frac{1}{2}\right)}_1\left(\omega \right)a^+_{in,1}\left(\omega \right)+C^{\left(-\frac{1}{2}\right)}_2}a^+_{in,2}\left(\omega \right){]\left.|0\right\rangle }_1{\left.|0\right\rangle }_2{\left.|-1/2\right\rangle }_c \nonumber \\
+{\alpha }_{1/2}\int^{\infty }_{-\infty }{d\omega [C^{\left(\frac{1}{2}\right)}_1\left(\omega \right)a^+_{in,2}\left(\omega \right)+C^{\left(\frac{1}{2}\right)}_2\left(\omega \right)}a^+_{in,1}\left(\omega \right){]\left.|0\right\rangle }_1{\left.|0\right\rangle }_2{\left.|1/2\right\rangle }_c\}.
    \end{array}
\label{eq15}
\end{equation}

As it is seen in Eq.(\ref{eq15}), the atom in state m=-1/2 preserves the initial quantum state of two mode photon state, while we have NOT gate for the photon state if m=1/2. This indicates the performance of CNOT gate in the case of monochromatic cavity. Bearing in mind that the atomic state (evaluation of the atomic amplitudes ${\alpha }_{-1/2}$ and ${\alpha }_{1/2}$) can be regulated by the controlling photon the photon-photon CNOT gate can be constructed using this approach.

\section{Possibilities of experimental implementation}

In order to construct the considered CNOT gate, we must satisfy the two matching conditions (\ref{eq13}). The first condition $\eta =\mu /2={\left|g\right|}^2/(\ \Delta -\eta )$ can be fulfilled  by choice of the photon-atom frequency detuning ${\Delta =\omega }_0-{\omega }_{ccw}$  for the given coupling constant of the circulating cavity light modes $\eta $.
In turns, the coupling constant $\eta $ is determined by the parameters of the fabricated Bragg grating (see Fig.1).
For  the fulfillment  of (\ref{eq13}), we can adjust the coupling coefficient $\kappa$ of the cavity with external waveguide by tuning the refractive index in the coupling region.
If the system parameters are properly chosen, we can efficiently realize the proposed CNOT gate by  following this scenario.
First, we use a controlling photon qubit for excitation of the controlling atom situated in the ring cavity.
The effective photon transfer to the atomic state can be realized by appropriated choice of the temporal photon mode.
In particular, this process can be performed for two-level or three-level lambda-type atom interacting with additional laser field on the adjoined atomic transition \cite{Gorshkov_2007}.
Moreover implementation of the considered CNOT gate on three-level atom could provide longer coherence time on the ground two levels that make it easier to perform.
Then, we load target photon qubit in the cavity that is not in resonance with the control atom at this stage.
Finally, we introduce CNOT gate operation cavity in resonance with its atom, and perform CNOT gate as it was described above.

Whispering-gallery mode optical cavity with high quality factor up to $Q = 10^{11}$ \cite{Savchenkov_2007} can be used to construct CNOT gate with enhanced coupling constant $\mu$. Besides controlling two-level atom (Rb for example) we suggest also quantum dot, nanoparticle with single active center attached to the surface of cavity or single atom or ion implanted in the cavity surface. Optical fiber-taper is widely used for coupling light field with cavity mode \cite{Knight_1997, Kippenberg_2002, Cai_2000}. For critical coupling very thin fiber is necessary with diameter $D <\lambda$ while such fibers are usually too unstable and can be damaged easily.
Recently integrated photonic waveguides were used for efficient coupling of atom with light \cite{Anderson_2018}.

In addition to this approach we suggest femtosecond laser written waveguide in fused silica plate \cite{Minnegaliev_2018, Chen_2013}.
Robust construction of this device is shown in Fig. \ref{fig:1}.
Laser written waveguide has small region where waveguide draw nearer (several $\mu$m) to the surface of fused silica plate.
This region can be used for coupling with whispering-gallery light modes through evanescent part of light field.
Coupling constant can be adjusted by changing the refractive index between cavity and plate or between waveguide and the surface of fused silica plate using short laser pulses.
Moreover the printed waveguide can be implemented in inorganic crystal doped by rare-earth ions \cite{Saglamyurek_2011, Zhong_2017, Minnegaliev_QE_2018} for on chip quantum storage of photonic qubits as it is depicted in Fig.\ref{fig:1}.
It is worth noting that the device can be efficiently connected with resonator quantum memory cells \cite{Andrianov_2015, Moiseev_2016, Simon_2010, Moiseev_2010, Sabooni_2013, Jobez_2014, Minnegaliev_QE_2018} for operation with large number of photonic qubits.

At the fabrication of waveguide, laser forms the shell of waveguide (two or more damage tracks) with decreased refractive-index, in addition this damage tracks produce stress field that leads to increasing of core refractive-index \cite{Chen_2013}.
It was shown that spectroscopic and coherency properties of rare-earth ions in materials perspective for implementation of quantum computer were not changed drastically \cite{Minnegaliev_2018, Corrielli_2016}.
Our measurements in waveguides produced in $^7LiYF_4:Er^{3+}$ show that erbium ions in these waveguides have coherent time sufficiently long  for quantum informatics  and waveguides with a high diameter ($\sim 100 \mu m$) have relatively low waveguide propagation losses (0.6 dB/cm).

\begin{figure}
  \includegraphics[width=1\textwidth]{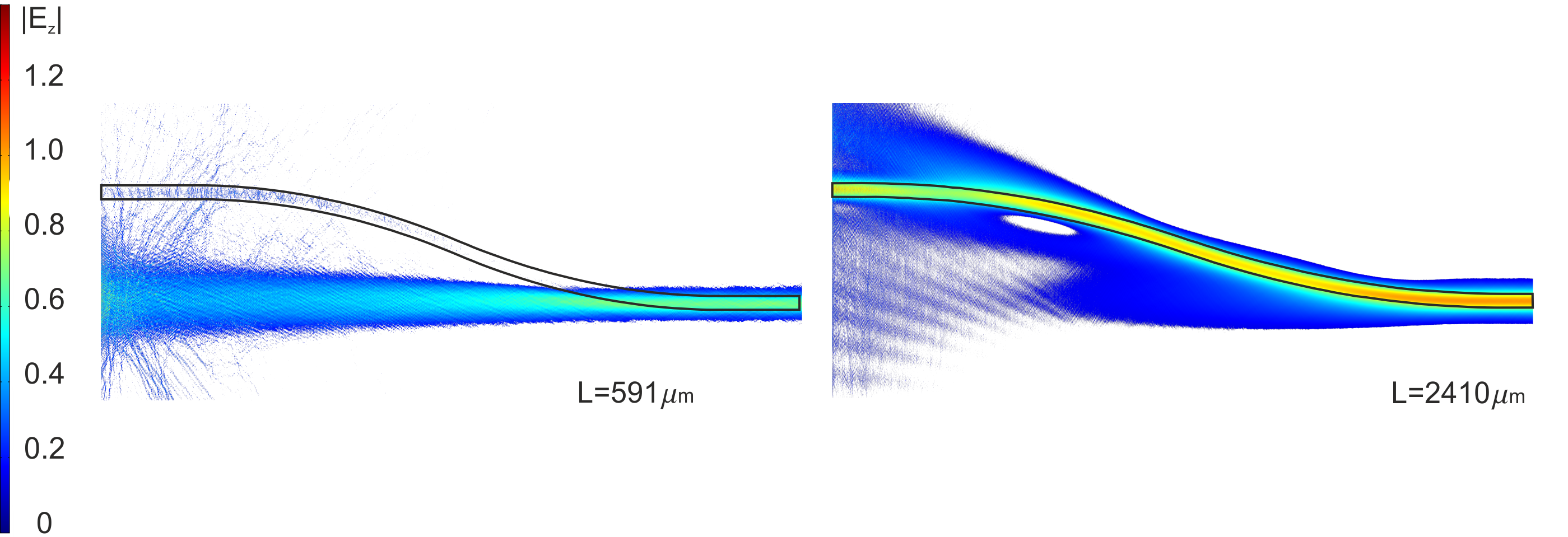}
\caption{Behavior of radiation with $E_y$ polarization in bent waveguide with core diameter 5 $\mu$m and waveguide length of L=0.591 mm and L= 2.41 mm.}
\label{fig:2}       
\end{figure}

However, with decreasing of waveguide diameter, these losses enhanced and the coherence time somewhat shortened.
Therefore we have investigated  another type of waveguides that are fabricated in $Y_2SiO_5$ (most popular material for quantum informatics applications \cite{Seri_2018}).
Laser beam produced a weak damage at the focal volume with relevant increase in refractive index and it is a position of waveguide core.
After fabrication of the waveguides, the difference of refractive indexes of the core with diameter $d_{core}$ and the coating is equal approximately $\triangle$n=0.001.
Such waveguides have demonstrated sufficiently broad radiation profile in the transverse dimension.
We have found that the transverse distribution of radiation is about 40 $\mu$m at the diameter of core equal $d_{core}= 5$ $\mu$m.
At the abrupt inclination of the waveguide the radiation get out of the waveguide as it is also shown in the numerical simulations depicted in  (Fig. \ref{fig:2}).

\begin{figure}
  \includegraphics[width=0.7\textwidth]{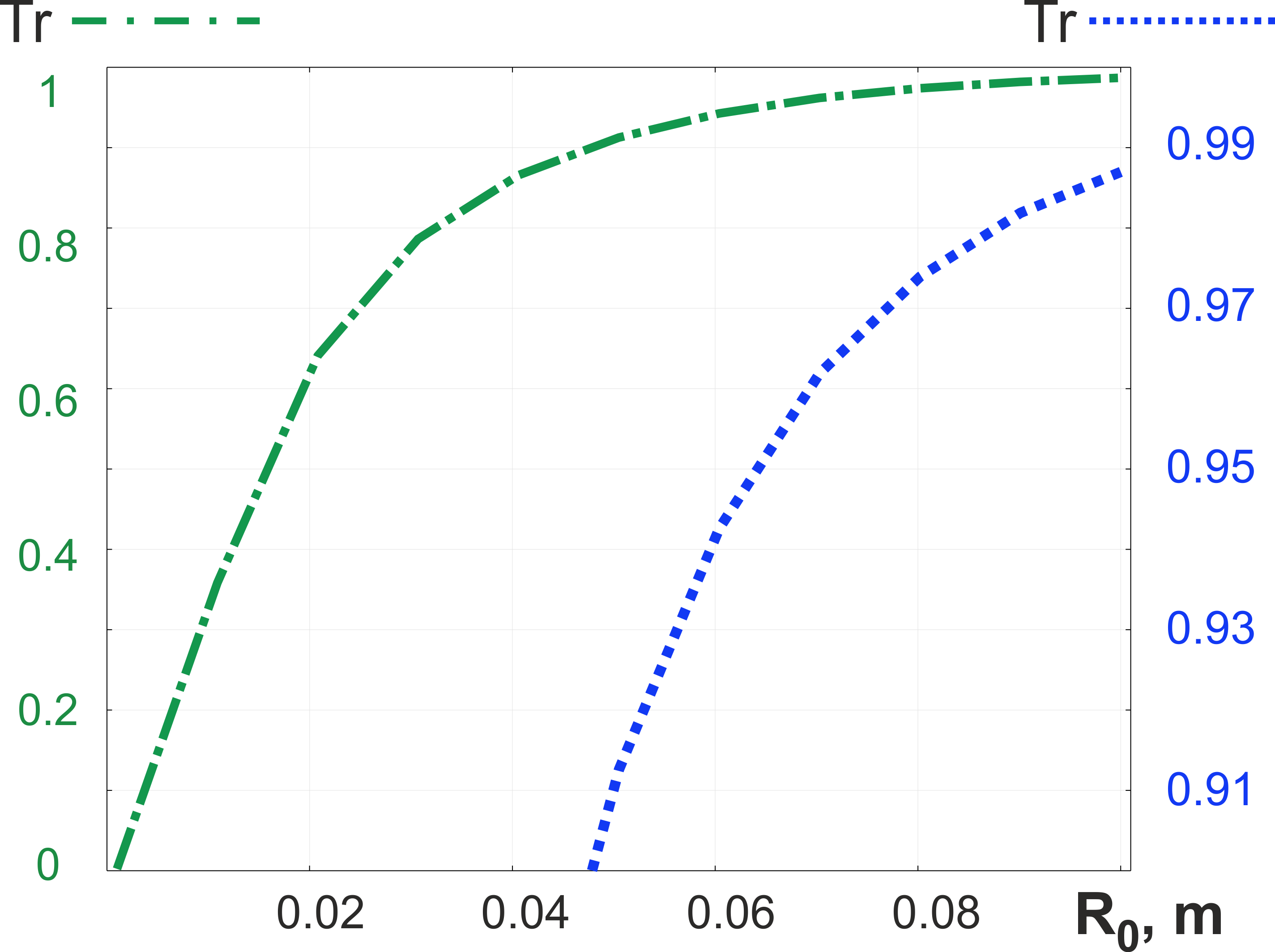}
\caption{The transmission Tr of light with $E_z$ polarization through the waveguide with core diameter 5 $\mu$m depending on the bending radius $R_0$ (the two curves are given in different scales).}
\label{fig:3}
\end{figure}

Therefore, the waveguide  should  have sufficiently large bending radius R.
The waveguide transmission can reach unity at the optimal parameters of bending, $d_{core}$ etc.
The light transmission through the  waveguide with diameter $d_{core}=5$ $\mu$m and 40 $\mu$m bending size is depicted in Fig.\ref{fig:3}.
For example, the transmission can reach $Tr=$97.5\%  in the operation zone at the radius R near 0.08 m that corresponds to the inclination angle 1.28 degrees.
In this case, the bending length will be about 3.6 mm.
As it is shown in Fig.\ref{fig:1}, a series of waveguide  with core size of 5 $\mu$m was created in a quartz glass with length 11.8 mm by using the femtosecond laser writing method.
The repetition rate of laser pulses creating waveguides was chosen 1 MHz, the pulse length - 400 fs, the average radiation power - 82 mW.
The laser output was frequency doubled to meet optimal focusing conditions with a Mitutoyo Plan APO 100X microscope objective \cite{Dyakonov_2016}.
The two bending zones have a  total length of 7.2 mm, which determines  the maximum length 4.6 mm of the interaction zone between the cavity and waveguide.
The creation of a curved waveguide was carried out at a depth of 55 $\mu$m with the subsequent rise of the waveguide core to the surface of the optical chip to a depth of 15 $\mu$m.
Such shape of the waveguides  ensured the steady propagation of light and its localization near the waveguide at a distance of 40 $\mu$m (as it was found by the numerical calculations).
After creating the waveguides, the ends of the optical chip were sanded to a depth of 150 $\mu$m and polished to provide the best input / output of the optical signal.
The radius of curvature we chose was 80mm, which provided a transfer efficiency in a double bending waveguide structure \ref{fig:1} of more than 90\%  in accordance with numerical calculation for 2 bend $(Tr)^2=95\%$ see Fig. \ref{fig:3}.

\begin{figure}
  \includegraphics[width=1\textwidth]{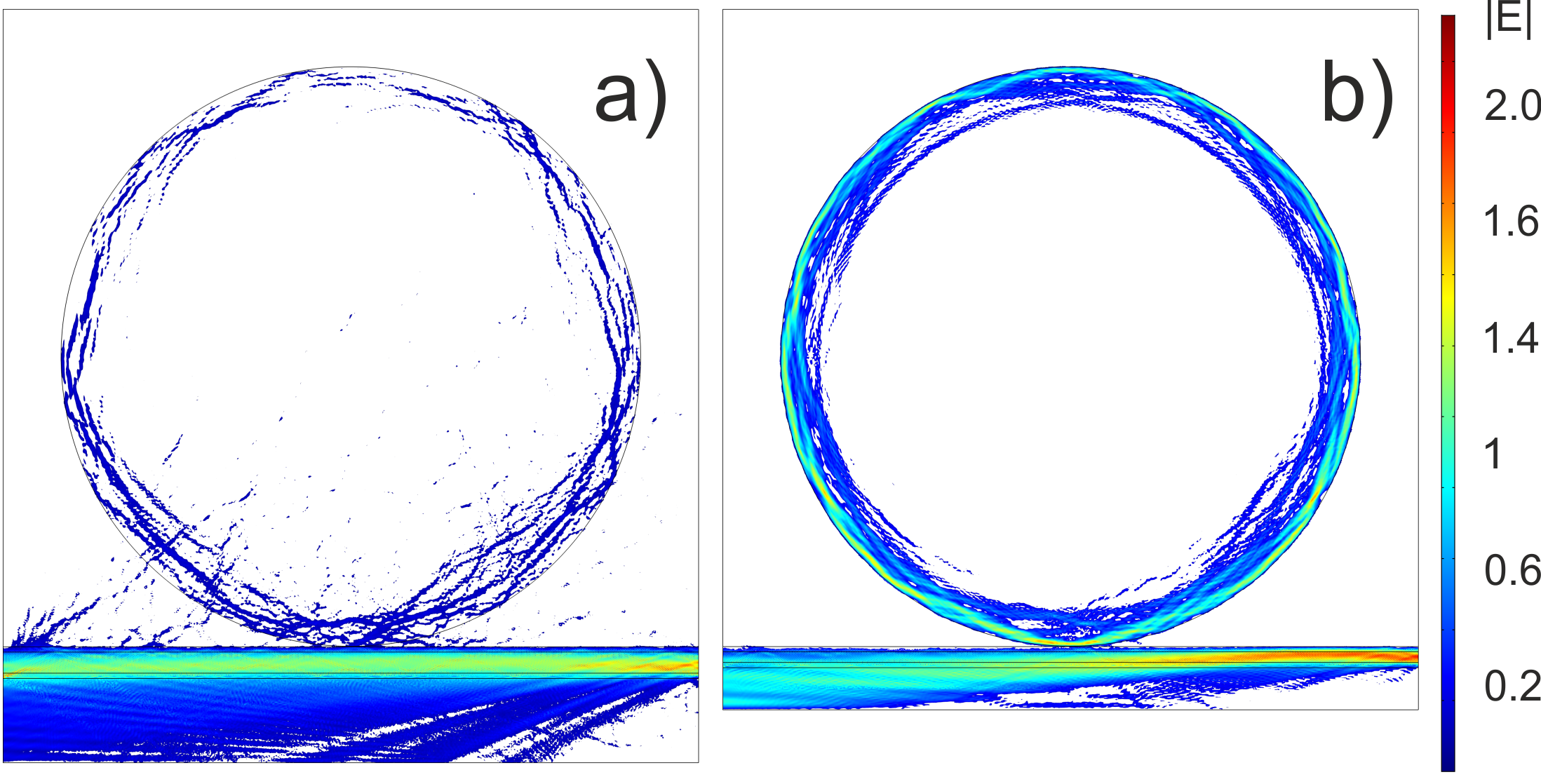}
\caption{ Nnmerical simulation of the  ring cavity excitation by the waveguide field (the waveguide mode is excited from the right side mode).
a) Diameter of the cavity is $D_{cavity}=110$ $\mu$m, wavelength is $\lambda=808 nm$; A) Diameter of the waveguide is $d_{core}=4$ $\mu$m.
b) Diameter of the waveguide is $d_{core}=2$ $\mu$m.
It is seen the enhancement of the ring cavity mode and out waveguide mode at the decreasing of waveguide diameter $d_{core}$ up to 2 $\mu$m and the field scatters partially in air.}
\label{fig:4}       
\end{figure}

In order to efficiently transfer radiation into the cavity, we need to find the appropriate parameters of the working area near the surface with the optimal depth of the waveguide and the length (see Fig.\ref{fig:4}).
To ensure the interaction of the optical field of the waveguide with the microcavity, the chip surface can be polished, which determines the actual depth of the waveguide area interacting with the microcavity.
We have numerically simulated the  radiation properties in such waveguide systems.
The light will reflect from the crystal-air interface  at the approach of radiation to the operative area (where ring cavity is located).
As a result, the maximum of  radiation distribution in transverse direction of non-symmetric waveguide will be shifted deeper in the crystal. This effect is due to the small refractive indexes difference.
This waveguide structure has an optimal distance for radiation transfer from the waveguide to the crystal-air interface $d_{surf}$ and then to resonator, correspondingly.
In this case, the initial diameter of the waveguide $d_{core}$ must be also diminished up to some optimal value.
For instance, the excitation of resonator (with the diameter $D_{cavity}=110$ $\mu$m and refractive index of crystal $n_{cavity}=1.47$) is shown in Fig.\ref{fig:4} for radiation wavelength $\lambda=808$ nm and polarization along vertical axis $E_y$ at the distance from waveguide to interface $d_{surf}=1$ $\mu$m.
Thus, performed calculations had shown that the effective excitation of resonator with diameter 110 $\mu$m is possible at the waveguide core diameter 2 $\mu$m and the waveguide bending radius 0.1 m.


\section{Conclusion }

We have proposed the photon-photon CNOT gate on counter propagating photon modes in ring cavity containing a two-level atom.
It was theoretically found that the CNOT gate can be realized at the optimal coupling constants of photon-atom interaction in the ring cavity and coupling constant of the photons propagating clockwise and counterclockwise.
It is shown that the CNOT gate can be realized deterministically with small number of operation steps in a simple optical scheme.
In our scheme, besides the proposed photon-photon CNOT gate, the single qubit gates can be also implemented by additional control of atomic state  which is necessary for the universal quantum computing.

We considered possible implementation of the proposed CNOT operation in on a chip waveguide scheme.
We have conducted a numerical simulation for such a scheme and experimental studies of the light propagation in the femtosecond laser written bending waveguide fabricated in quartz glass.
The optimal parameters of the waveguide systems were found for 90\% efficiency of waveguided light transmission to the cavity area.
Thus, the performed studies demonstrated possible ways for the on a chip implementation of the proposed photon-photon CNOT gate.


\bibliographystyle{spphys}       
\bibliography{biblio}   

\end{document}